\documentclass[conference]{IEEEtran}
\IEEEoverridecommandlockouts

\usepackage{graphicx}
\usepackage{textcomp}
\usepackage{xcolor}
\usepackage{breqn}
\usepackage{algorithmic}
\usepackage[numbers,sort&compress]{natbib}
\usepackage{array}
\usepackage{booktabs}
\usepackage{times}
\usepackage{setspace}
\usepackage{etoolbox}
\usepackage[font=small,skip=2pt]{caption}

\begin{document}
\setstretch{0.85} %% This paragraph or section will have tighter line spacing.

%% Metadata
\title{Trust Dynamics in Cryptocurrency Markets: Centralized vs. Decentralized Exchanges}

\author{
\IEEEauthorblockN{Xintong Wu\textsuperscript{$\dagger$}}
\IEEEauthorblockA{
\textit{University of Pennsylvania}\\ 
New York, NY, USA
}
\and
\IEEEauthorblockN{Wanlin Deng\textsuperscript{$\dagger$} and Yutong Quan}
\IEEEauthorblockA{
\textit{Columbia University}\\ 
New York, NY, USA
}
\and
\IEEEauthorblockN{Lin William Cong\textsuperscript{$\ddagger$}}
\IEEEauthorblockA{
\textit{Nanyang Technological University}\\ 
Singapore
}
\and 
\IEEEauthorblockN{Luyao Zhang\textsuperscript{*}}
\IEEEauthorblockA{
\textit{Duke Kunshan University}\\ 
Suzhou, Jiangsu, China}

\thanks{\textsuperscript{*}Corresponding author: Luyao Zhang (lz183@duke.edu), Digital Innovation Research Center and Social Science Division, Duke Kunshan University. Address: Duke Avenue No.8, Kunshan, Suzhou, Jiangsu, China, 215316. \textbf{Acknowledgments}: Zhang acknowledges the support from the National Science Foundation China (NSFC) under the project titled "Trust Mechanism Design on Blockchain: An Interdisciplinary Approach of Game Theory, Reinforcement Learning, and Human-AI Interactions" (Grant No. 12201266). Quan, Wu, and Deng are grateful for the support from the Summer Research Scholar Program and the Wang-Cai Biochemistry Lab Donation Fund at Duke Kunshan University, supervised by Prof. Zhang; they also appreciate the support from the Center for Study of Contemporary China (CSCC) Undergraduate Research Grant, with Prof. Zhang as the principal investigator. Finally, Cong acknowledges the generous support from Ripple's University Blockchain Research Initiative. \textsuperscript{$\ddagger$}Lin William Cong is also with People's Bank of China School of Finance (Special-Term), Tsinghua University. \textsuperscript{$\dagger$} joint first authors.}
}

\maketitle

%% Abstract
\begin{abstract}
Trust mechanisms diverge between centralized and decentralized exchanges, representing distinct sociotechnical governance paradigms. However, quantifying trust dynamics and their redistribution between these architectures remains empirically challenging, limiting understanding of how institutional shocks affect market behavior. The FTX collapse offers a natural experiment to bridge this gap. Through an interdisciplinary approach combining causal inference and computational text analysis, we find significant price declines and capital reallocation from centralized to decentralized exchanges following the event. While sentiment metrics showed no sharp discontinuities, topic modeling and network analysis of Discord communities reveal that seasonal holiday discourse obscured underlying trust concerns in centralized exchange forums. These findings underscore the fragility of institutional trust architectures and demonstrate how mixed methods can illuminate behavioral patterns during systemic crises, offering insights for exchange risk management and regulatory assessment.
\end{abstract}

\begin{IEEEkeywords}
Centralized exchanges, Decentralized exchanges, Causal inference, Natural language processing, Topic modeling,  Financial markets, Blockchain, Cryptocurrency exchanges
\end{IEEEkeywords}

\section{Introduction}
\begin{figure}[!t]
\centering
\includegraphics[width=0.85\linewidth]{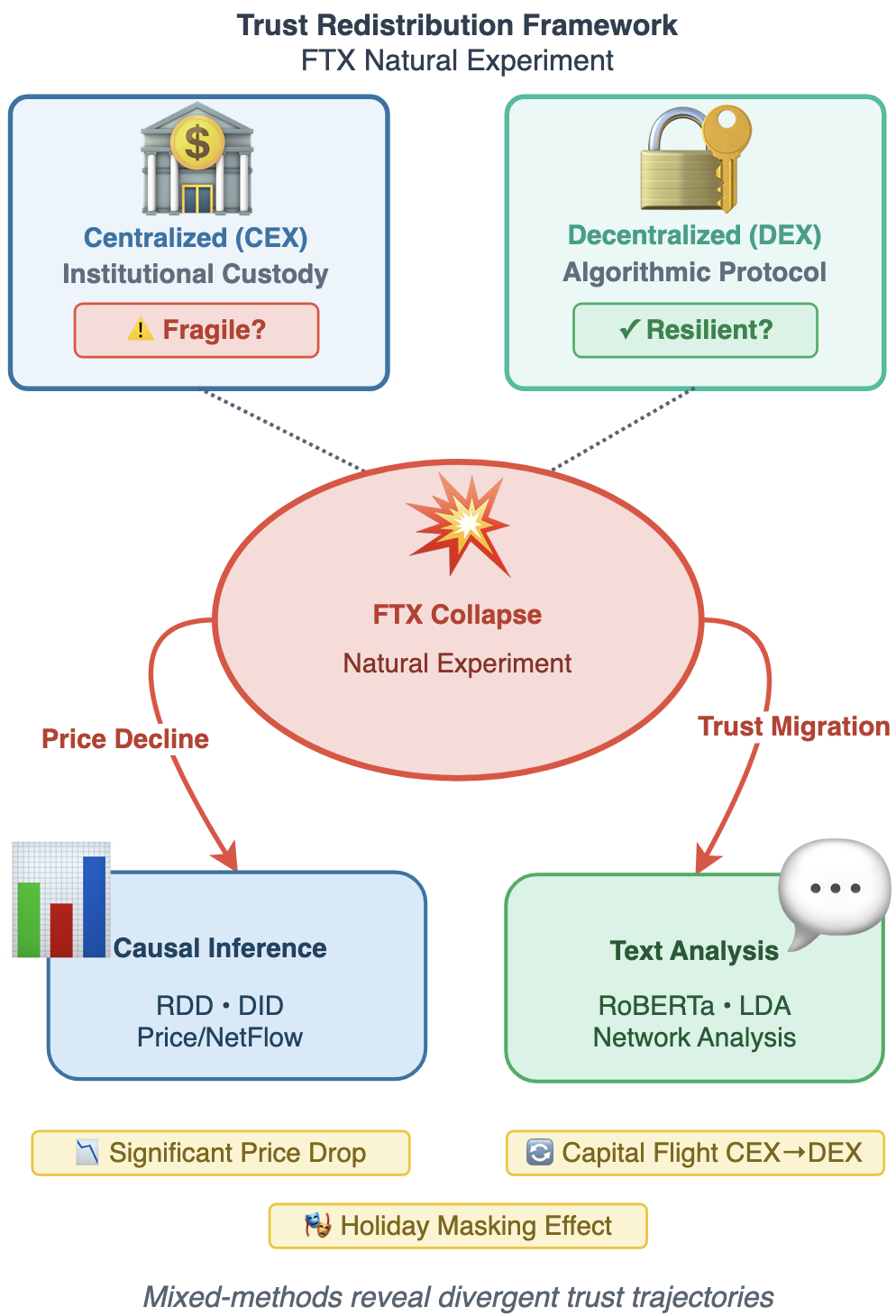}
\caption{Trust Redistribution Framework: The FTX collapse as a natural experiment contrasting institutional custody (CEX) and algorithmic protocol (DEX) architectures, analyzed through mixed causal inference and computational text methods.}
\label{fig:framework}
\end{figure}

Trust, requiring the careful alignment of technical capabilities with ethical responsibility and institutional governance \citep{10.1145/3789256}, serves as a fundamental element of human society and plays a critical role in the digital economy. Blockchain technology exemplifies this sociotechnical nature, representing a paradigm shift from institutional trust mechanisms to algorithmic and community-based alternatives. However, the trustworthiness of these systems---whether centralized exchanges (CEXs) or decentralized exchanges (DEXs)\cite{hagele2024centralized}---depends on a complex interplay of technical performance \citep{augusto2024sok,liu2022deciphering,chemaya2023uniswap,yan2024analyzing,10704461,fu2024quantifying}, governance structures \citep{ao2023decentralized,zhang2023blockchain,quan2024decoding,xiao2023centralized}, and ethical alignment \citep{fu2023ai,zhang2023mechanics}. Numerous studies have shown that trust influences cryptocurrency market development and investor behavior \citep{park2022impact, rehman_salah_damiani_svetinovic_2019, arli_van_esch_bakpayev_laurence_2020, marella_upreti_merikivi_tuunainen_2020}. 

Despite its importance, trust remains difficult to measure, limiting the study of its effects on market dynamics. These measurement challenges are particularly acute across cryptocurrency exchanges, where two divergent trust architectures prevail: centralized exchanges (CEXs) relying on institutional custody, and decentralized exchanges (DEXs) utilizing algorithmic protocols. CEXs, operated by a central organization, require users to trust the exchange with their assets in shared wallets. Transactions on CEXs occur off the blockchain, relying on the exchange to manage trades. Trust in CEXs is typically based on reputation, user base, or endorsements. However, events like the collapse of FTX highlight the trust issues arising from regulatory breaches \citep{victor2021detecting}. In contrast, DEXs utilize smart contracts or peer-to-peer networks for non-custodial trading. The two main DEX implementations are limit order books (e.g., EtherDelta) and automated market makers (e.g., PancakeSwap) \citep{hagele2024centralized}. DEXs eliminate the need for trust in a central entity by relying on predetermined and automated transaction rules. While this resolves traditional trust issues, challenges such as technical complexity, lack of regulation, and security threats introduce new trust considerations, including the potential for market manipulation \citep{10190515}.

The collapse of FTX serves as a prominent example of a risk event that has profoundly impacted the cryptocurrency market. On November 11, 2022, FTX Trading announced voluntary bankruptcy proceedings. The misappropriation of user assets and governance failures triggered significant erosion of trust in centralized institutions, with ripple effects across the broader market \citep{conlon2022collapse}. This pivotal event raised questions about the resilience of centralized trust systems and offers a unique natural experiment to examine how sudden trust erosion propagates through cryptocurrency markets, affecting user behavior, token valuation, trading flows, and sentiment dynamics. Figure \ref{fig:framework} illustrates this natural experiment framework, contrasting institutional and algorithmic trust architectures while mapping the capital migration and analytical methods employed in this study. Inspired by the insights of Park et al. \citep{park2022impact}, our study delves into the dynamics of the cryptocurrency market and the mechanisms underlying user trust. Employing causal inference methods, including Regression Discontinuity Design (RDD) and Difference-in-Differences (DID), this research reveals significant declines in WETH prices and NetFlow from CEXs to DEXs, signaling a measurable transfer of trust. Additionally, natural language processing methods, including topic modeling and sentiment analysis, uncover the complexities of user responses, highlighting shifts from functional discussions to emotional fragmentation in Binance's community, while Uniswap's sentiment exhibits a gradual upward trend. Despite data limitations and external influences, the findings underscore the intricate interplay between trust, sentiment, and market behavior in the cryptocurrency ecosystem.

Through interdisciplinary research integrating Trust Theory \citep{10.1145/3789256}, our objective is to enhance the understanding of trust mechanisms---characterized by \textit{transparency}, \textit{accountability}, and \textit{reliability}---within crypto-asset markets. The FTX collapse constitutes a breach of these core trust principles, providing a natural experiment to examine how such violations propagate through market prices, platform choice, and user sentiment. Within this framework, our primary research questions are:
\begin{itemize}
    \item \textbf{RQ1:} How does the breach of trust principles (transparency, accountability, reliability) during the FTX collapse affect prices across the crypto market?
    \item \textbf{RQ2:} How does the erosion of \textit{institutional trust} in centralized exchanges (CEX) versus \textit{distributed trust} in decentralized exchanges (DEX) drive platform migration?
    \item \textbf{RQ3:} How does user sentiment regarding trust violations evolve on social media platforms (e.g., Discord) during the collapse?
\end{itemize}

\subsection{Hypotheses}
The FTX collapse provides a natural experiment to test three hypotheses regarding trust migration: 

\begin{itemize}
   \item \textbf{H1 (Market-wide Price Impact):} Breaches of \textit{transparency} and \textit{accountability} at FTX trigger systemic trust erosion, extending fear to the broader ecosystem and triggering price declines---drawing on documented evidence that FTX triggered systemic market contagion \citep{conlon2022collapse}. 
    \item\textbf{H2 (Inter-Platform Fund Migration):} Users withdraw \textit{institutional trust} from CEXs (generalizing FTX's governance failures) and migrate toward \textit{distributed trust} systems (DEXs), prompting NetFlow shifts---consistent with documented post-collapse migration patterns \citep{fu2023ftx}. 
    \item\textbf{H3 (Divergent Community Sentiment):} Overconfidence amplifies \textit{trust redistribution}, wherein Binance suffers categorical association with FTX's trust breaches while Uniswap gains from perceived distributed reliability---grounded in behavioral finance theory \citep{barberis2018psychology}.
\end{itemize}
We evaluate H1 and H2 via RDD and DID analyses of WETH prices and NetFlow dynamics, and H3 through sentiment analysis of Discord communities. These hypotheses bridge behavioral finance theories with blockchain analytics to quantify the socio-technical dimensions of trust in digital economies.

\textbf{Data and Code Availability Statement}: The dataset and code used in this study are openly available on GitHub\footnote{\texttt{https://github.com/Xintong1122/Event\_Study}}. The repository includes all datasets used for analysis, as well as the scripts for data processing and statistical modeling.

\section{Data and Method}

To capture the daily movement of funds between exchange types, we calculate net platform flow, denoted by the flux variable $\Phi$, as the difference between centralized and decentralized exchange inflows:
\begin{equation}
    \Phi = \mathcal{I}_{\text{CEX}} - \mathcal{I}_{\text{DEX}}
    \label{eq:netflow}
\end{equation}
where $\mathcal{I}_{\text{CEX}}$ represents the aggregate daily inflow to centralized exchanges (Coinbase, OKX, Binance) and $\mathcal{I}_{\text{DEX}}$ represents the aggregate daily inflow to decentralized exchanges (Uniswap, Sushiswap, Balancer). The resulting flux $\Phi$ (termed \textit{NetFlow} in the empirical analysis) indicates the directional movement of capital: positive values ($\Phi > 0$) signify capital concentration toward centralized platforms, interpreted as heightened trust in custodial institutions, while negative values ($\Phi < 0$) reflect capital migration toward decentralized protocols, signaling trust decentralization.

Complete variable definitions, empirical ranges, temporal coverage specifications, validation procedures, and anonymization protocols are documented in Appendix~\ref{app:data} and Table~\ref{tab:data_description}.

We evaluate the causal impact of the FTX collapse on WETH prices, NetFlow, and user sentiment using Regression Discontinuity Design (RDD) \citep{trochim1990regression, hahn2001identification}, Difference-in-Differences (DID) \citep{abadie2005semiparametric, callaway2021difference}, Word Frequency Analysis \citep{baayen2012word}, Latent Dirichlet Allocation (LDA) \citep{blei2003latent, dyer2017evolution}, and Network Analysis \citep{doncheva2018cytoscape}. Detailed methodological specifications appear in Appendix~\ref{app:method}.

\section{Results and Discussion}

\subsection{Price and NetFlow Dynamics}
\subsubsection{WETH Price}
\begin{figure}[ht!]
    \centering
    \includegraphics[width=0.9\columnwidth]{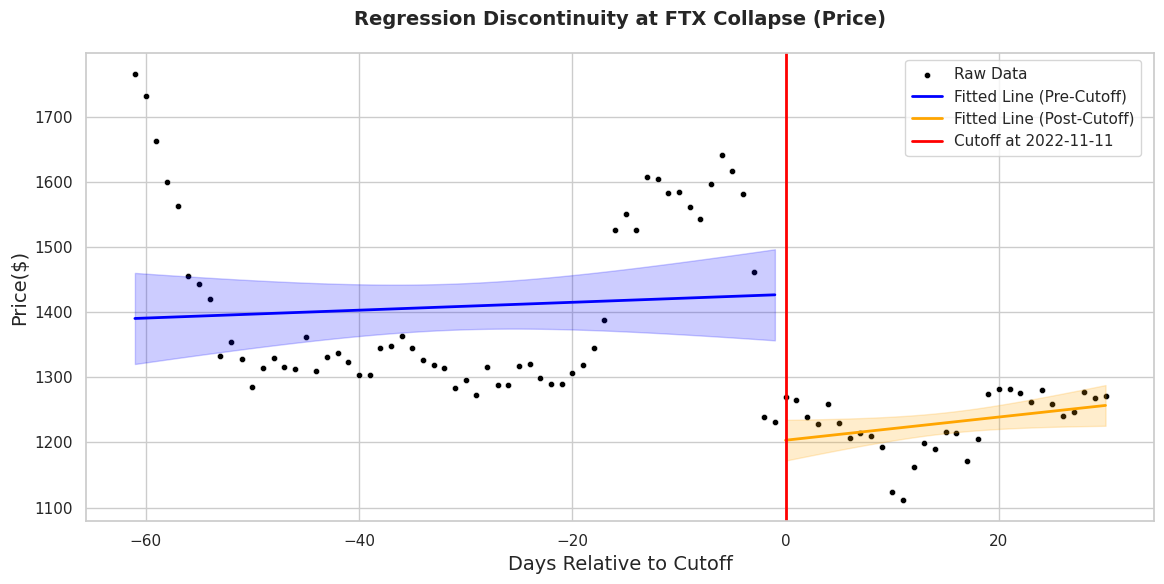}
    \caption{Regression Discontinuity Design for Price}
    \label{fig:price_rdd}
\end{figure}

The RDD test for WETH prices indicates a negative discontinuity of $-223.82$. at the threshold. The confidence interval excludes zero, demonstrating statistically significant price declines following the FTX collapse (Figure \ref{fig:price_rdd}). Complete RDD estimates for all outcomes are reported in Table~\ref{tab:rdd_results} (Appendix~\ref{app:results}).  These findings are consistent with our hypothesis that exchange failure eroded market confidence, contributing to WETH depreciation.

\begin{figure}[ht!]
    \centering
    \includegraphics[width=0.9\columnwidth]{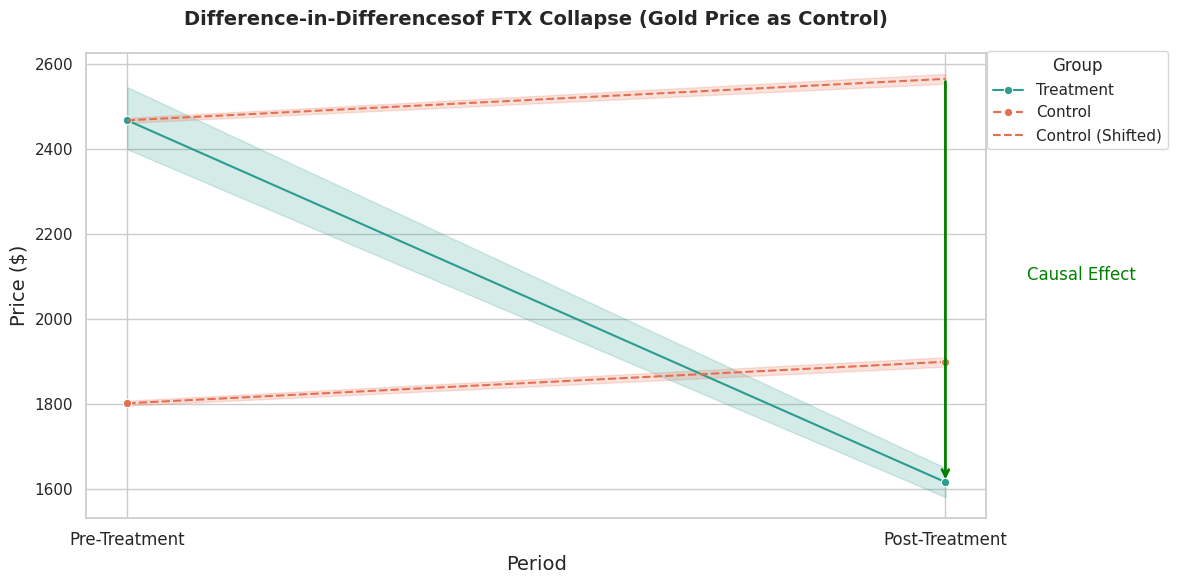}
    \caption{Difference-in-Differences for WETH Prices (Gold as Control)}
    \label{fig:price_did}
\end{figure}

The DID analysis reveals a treatment effect of $ 950.03$ with a 95\% confidence interval of $[-950.13,- 949.93]$. showing significant price divergence from gold benchmarks (Figure \ref{fig:price_did}). This differential trajectory confirms the event's specific impact on cryptocurrency valuations beyond broader market trends.

\subsubsection{WETH NetFlow}
\begin{figure}[ht!]
    \centering
    \includegraphics[width=0.9\columnwidth]{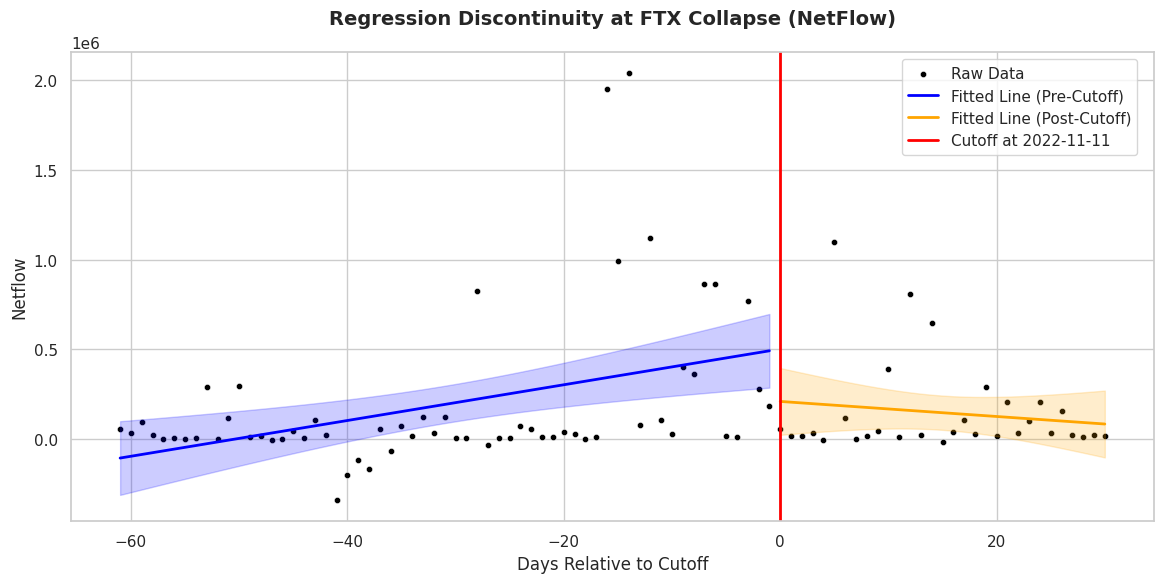}
    \caption{Regression Discontinuity Design for NetFlow}
    \label{fig:netflow_rdd}
\end{figure}

NetFlow analysis shows a discontinuity at the threshold of $-291,928.65$, with a confidence interval that does not include 0. The entirely negative confidence interval confirms a statistically significant decline in NetFlow from CEX to DEX after the FTX collapse (Figure \ref{fig:netflow_rdd}). This evidence supports our hypothesis of trust migration from centralized to decentralized platforms following exchange failures.

\subsection{Community Sentiment and Topic Analysis}
\subsubsection{Binance and Uniswap Community Sentiment}
\begin{figure}[ht!]
    \centering
    \includegraphics[width=0.9\columnwidth]{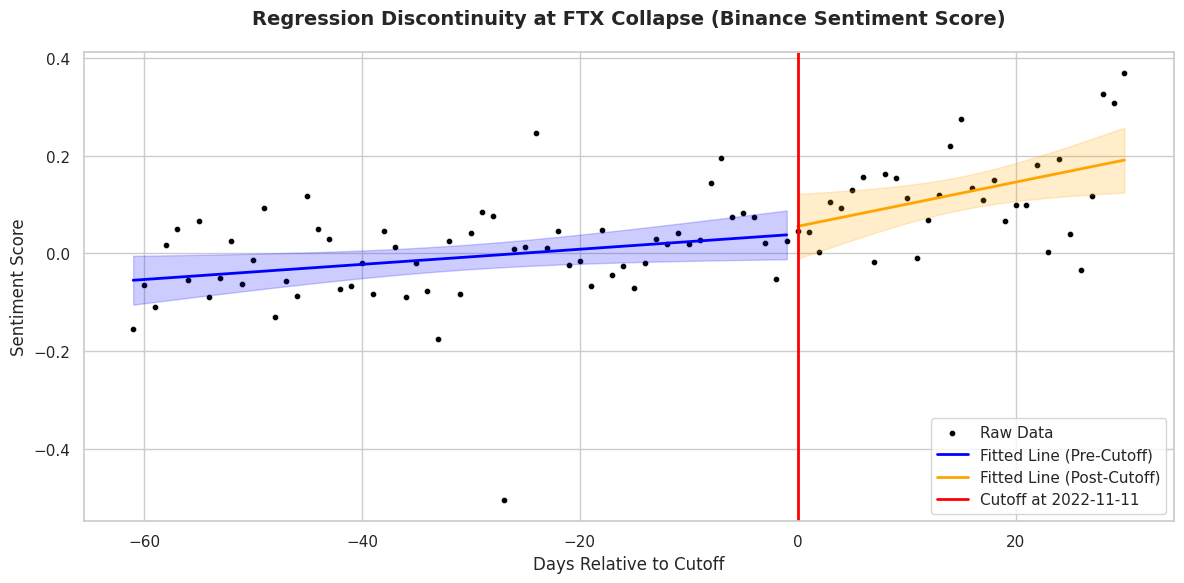}
    \caption{Regression Discontinuity Design for Binance Sentiment Score}
    \label{fig:senti_b_rdd}
\end{figure}

The RDD analysis reveals distinct sentiment dynamics in the Binance and Uniswap communities following the FTX collapse. For Binance, the sentiment shift showed a discontinuity at the threshold of $0.02$, with a 95\% confidence interval of $[-0.08, 0.12]$. While positive directional changes appear, the confidence interval spanning zero suggests no statistically reliable evidence of sentiment deterioration (Figure \ref{fig:senti_b_rdd}). This finding suggests that the FTX collapse did not lead to a notable decline in sentiment within the Binance community, contrary to our hypothesis.

\begin{figure}[ht!]
    \centering
    \includegraphics[width=0.9\columnwidth]{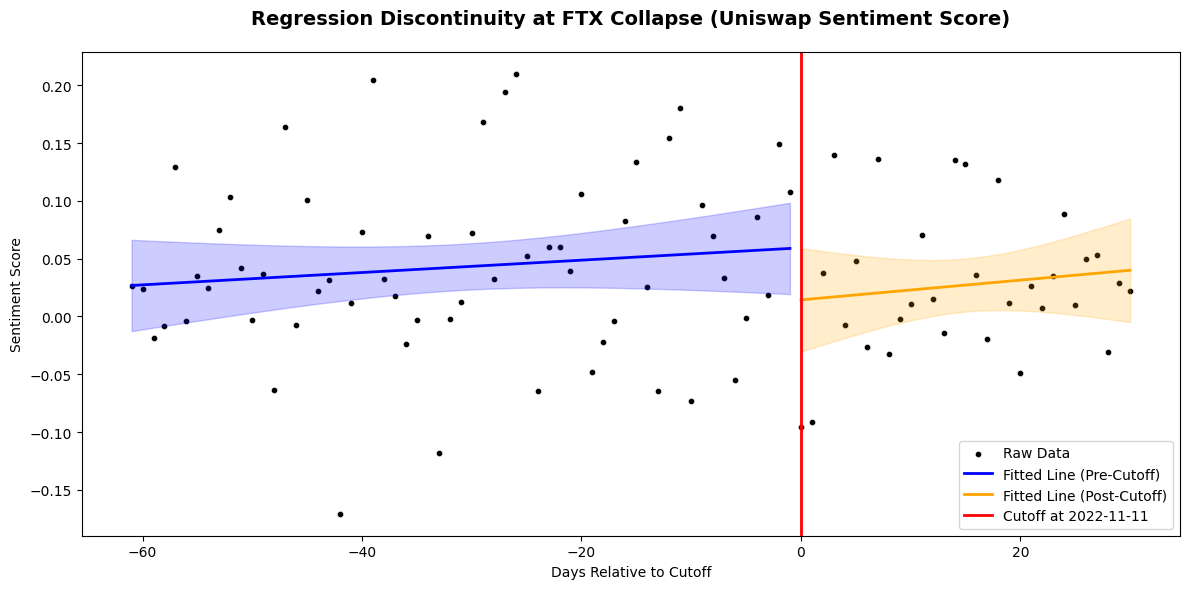}
    \caption{Regression Discontinuity Design for Uniswap Sentiment Score}
    \label{fig:senti_u_rdd}
\end{figure}

The RDD analysis of Uniswap sentiment demonstrated a discontinuity of $-0.05$, with a 95\% confidence interval of $[-0.15, 0.05]$. Although this result is not statistically significant, it does not directly contradict the hypothesis. Specifically, an upward trend was observed in the sentiment scores of Uniswap following an initial decline (Figure \ref{fig:senti_u_rdd}). This suggests that the FTX collapse may have eventually driven users to favor DEX platforms like Uniswap, aligning with the hypothesis that trust shifted from CEX to DEX after the event.

\subsubsection{Binance Further Topic Analysis}
Given that regarding Binance sentiment, these results contradict our hypothesis, we conducted a more detailed analysis of the community's discussion content to understand the underlying dynamics better. The study focused on word frequency, topic modeling using Latent Dirichlet Allocation (LDA), and network analysis to uncover the nuances of user sentiment and the impact of the FTX collapse on discussion patterns.

\begin{figure}[ht!]
    \centering
    \includegraphics[width=0.85\columnwidth]{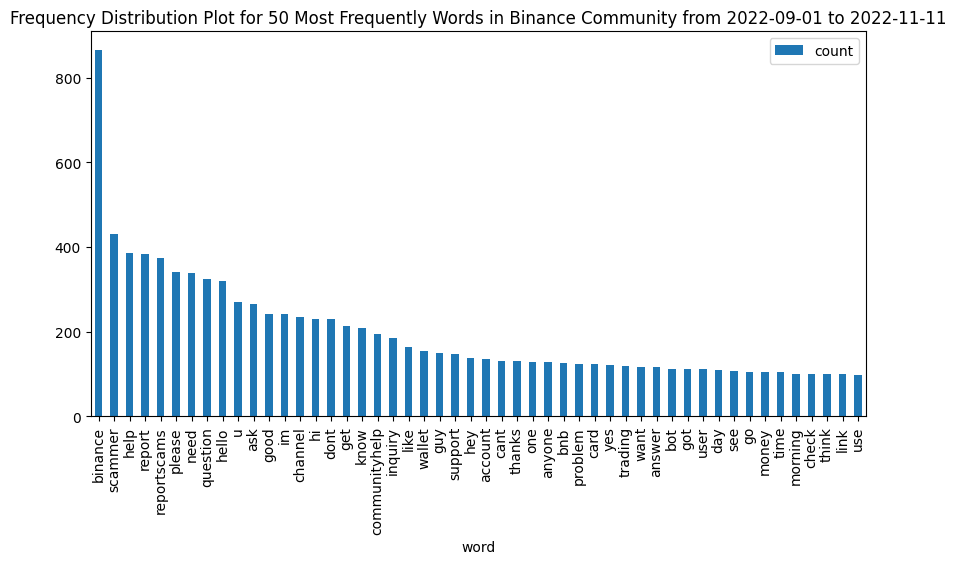}
    \caption{Top 50 Words in Binance Community Pre-FTX Collapse}
    \label{fig:word_freq1}
\end{figure}

\begin{figure}[ht!]
    \centering
    \includegraphics[width=\columnwidth]{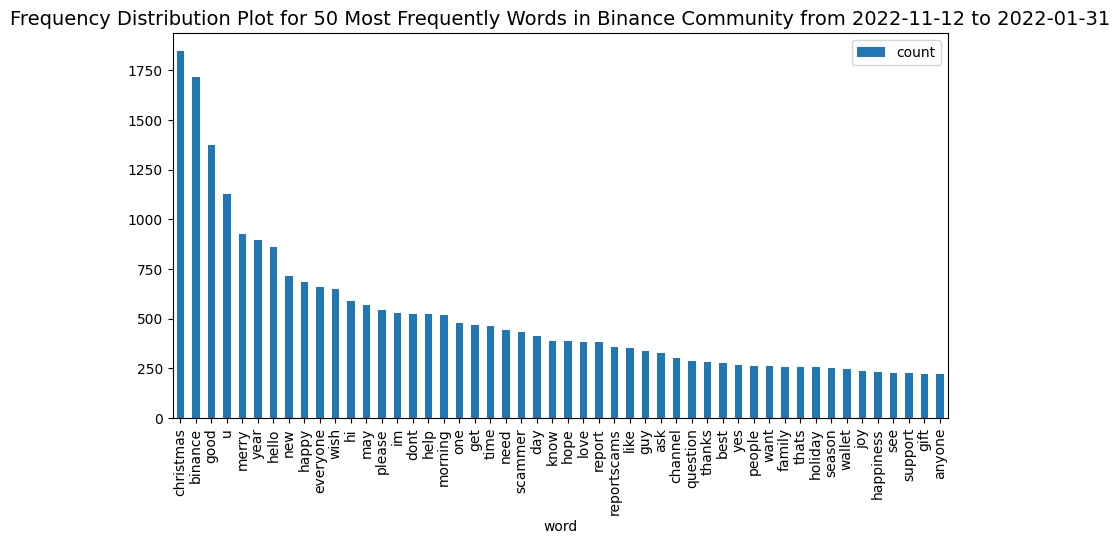}
    \caption{Top 50 Words in Binance Community Post-FTX Collapse}
    \label{fig:word_freq2}
\end{figure}

Word frequency analysis reveals a marked shift in community vocabulary pre- and post-collapse (Figures \ref{fig:word_freq1} and \ref{fig:word_freq2}). Pre-FTX, discourse centers on functional and support-related terms, with \textit{binance,'' scammer,'' help,'' report,''} and \textit{wallet''} dominating the frequency distribution (Fig. \ref{fig:word_freq1}). Post-FTX, the lexicon shifts dramatically toward holiday-related expressions; \textit{christmas,'' merry,'' year,'' good,''} and \textit{happy''} surge to the highest frequencies (Fig. \ref{fig:word_freq2}), suggesting that seasonal sentiment potentially masked negative reactions to the FTX failure.

\begin{figure}[ht!]
    \centering
    \includegraphics[width=\columnwidth]{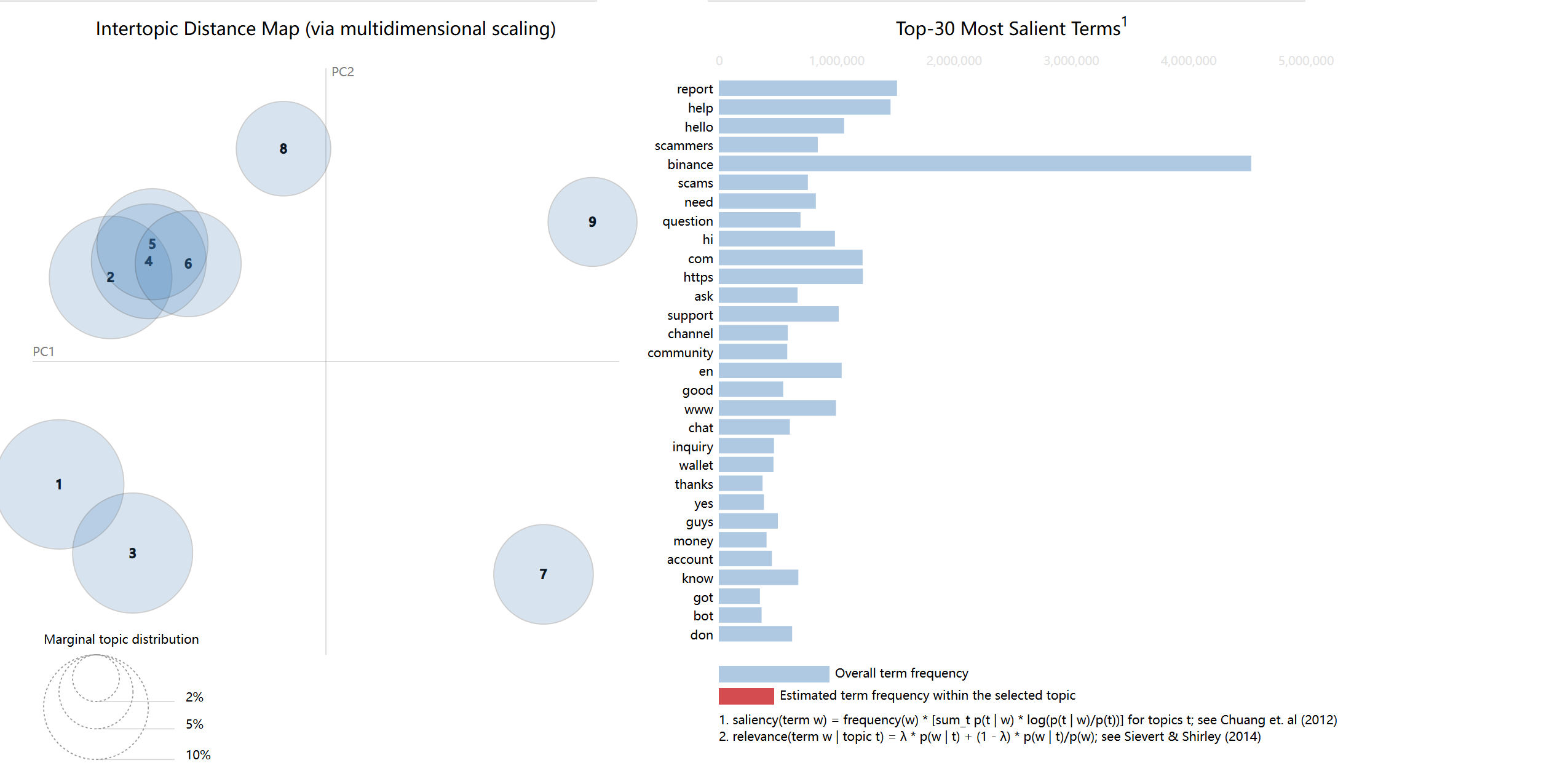}
    \caption{Intertopic Distance Map and Top-30 Terms Pre-FTX Collapse}
    \label{fig:lda1}
\end{figure}

\begin{figure}[ht!]
    \centering
    \includegraphics[width=\columnwidth]{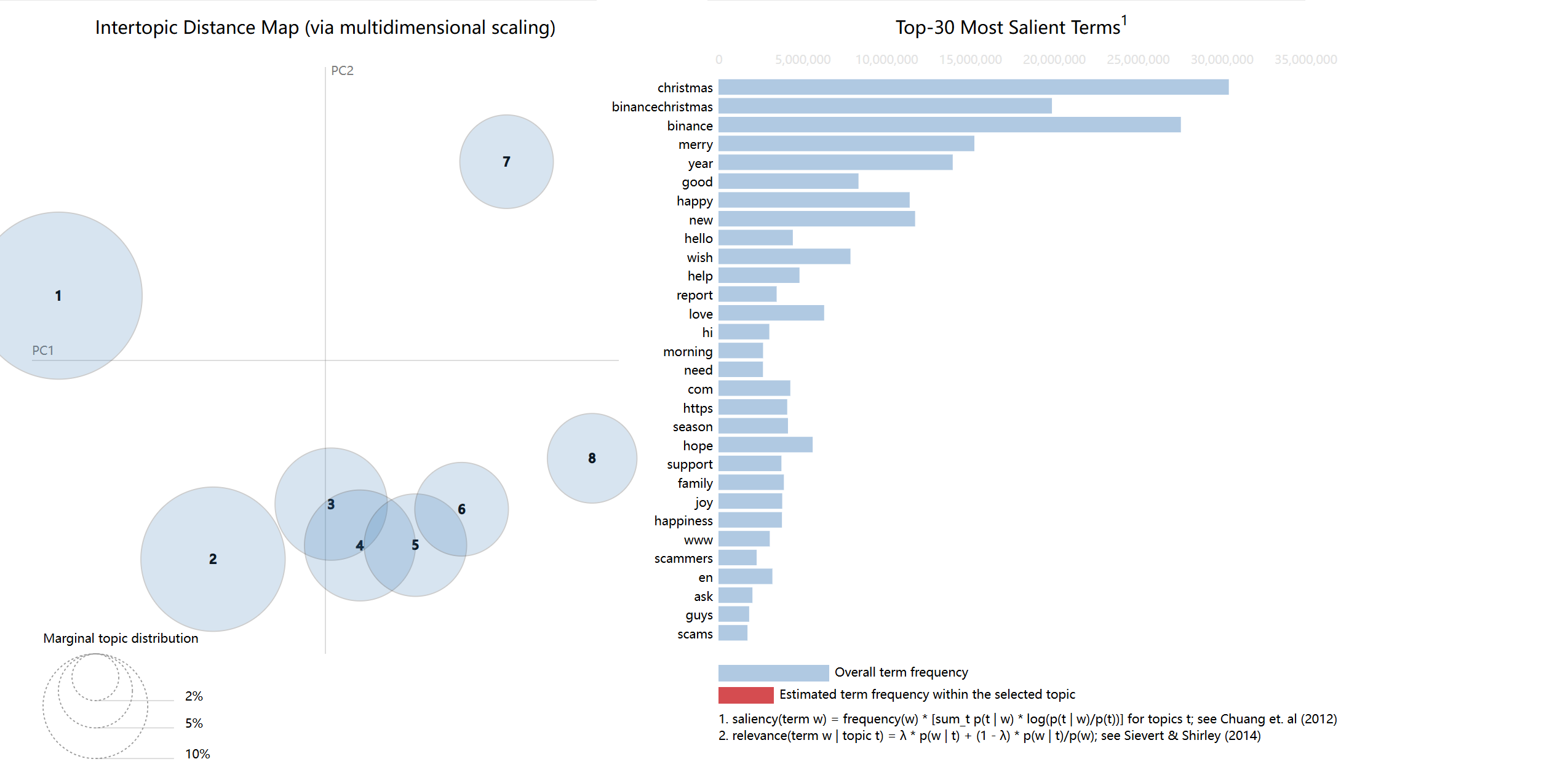}
    \caption{Intertopic Distance Map and Top-30 Post-FTX Collapse}
    \label{fig:lda2}
\end{figure}

Topic modeling via LDA corroborates this thematic displacement (Figures \ref{fig:lda1} and \ref{fig:lda2}). Pre-collapse, discussions cluster tightly around operational support and security reporting, evidenced by salient terms including \textit{report,'' help,'' scammers,'' wallet,''} and \textit{account''} (Fig. \ref{fig:lda1}). Post-collapse, the topic distribution fragments; the discourse is overwhelmingly dominated by holiday-related topics, characterized by high-saliency terms such as \textit{christmas,''} \textit{binancechristmas,''} \textit{merry,''} \textit{wish,''} and \textit{holiday''} (Fig. \ref{fig:lda2}). This shift from cohesive functional discussions to fragmented, seasonally-focused discourse illustrates how the collapse coincided with---and was potentially obscured by---surging holiday sentiment, despite underlying concerns about platform trust. Detailed topic-term mappings are provided in Table~\ref{tab:topic_words} (Appendix~\ref{app:results}).

\begin{figure}[ht!]
    \centering
    \includegraphics[width=0.6\columnwidth]{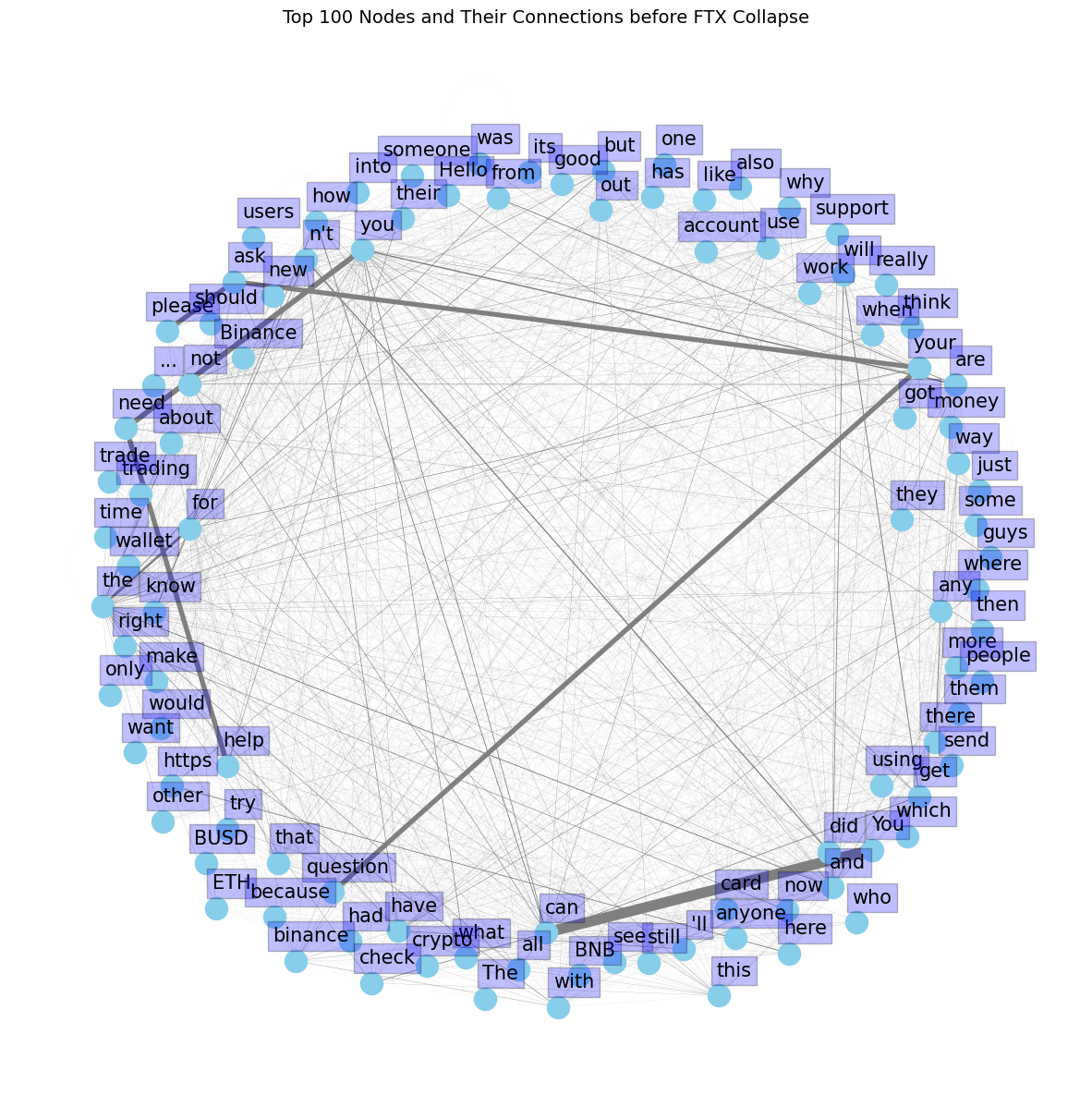}
    \caption{Network Analysis of Top 100 Nodes Pre-FTX Collapse}
    \label{fig:net1}
\end{figure}

\begin{figure}[ht!]
    \centering
    \includegraphics[width=0.6\columnwidth]{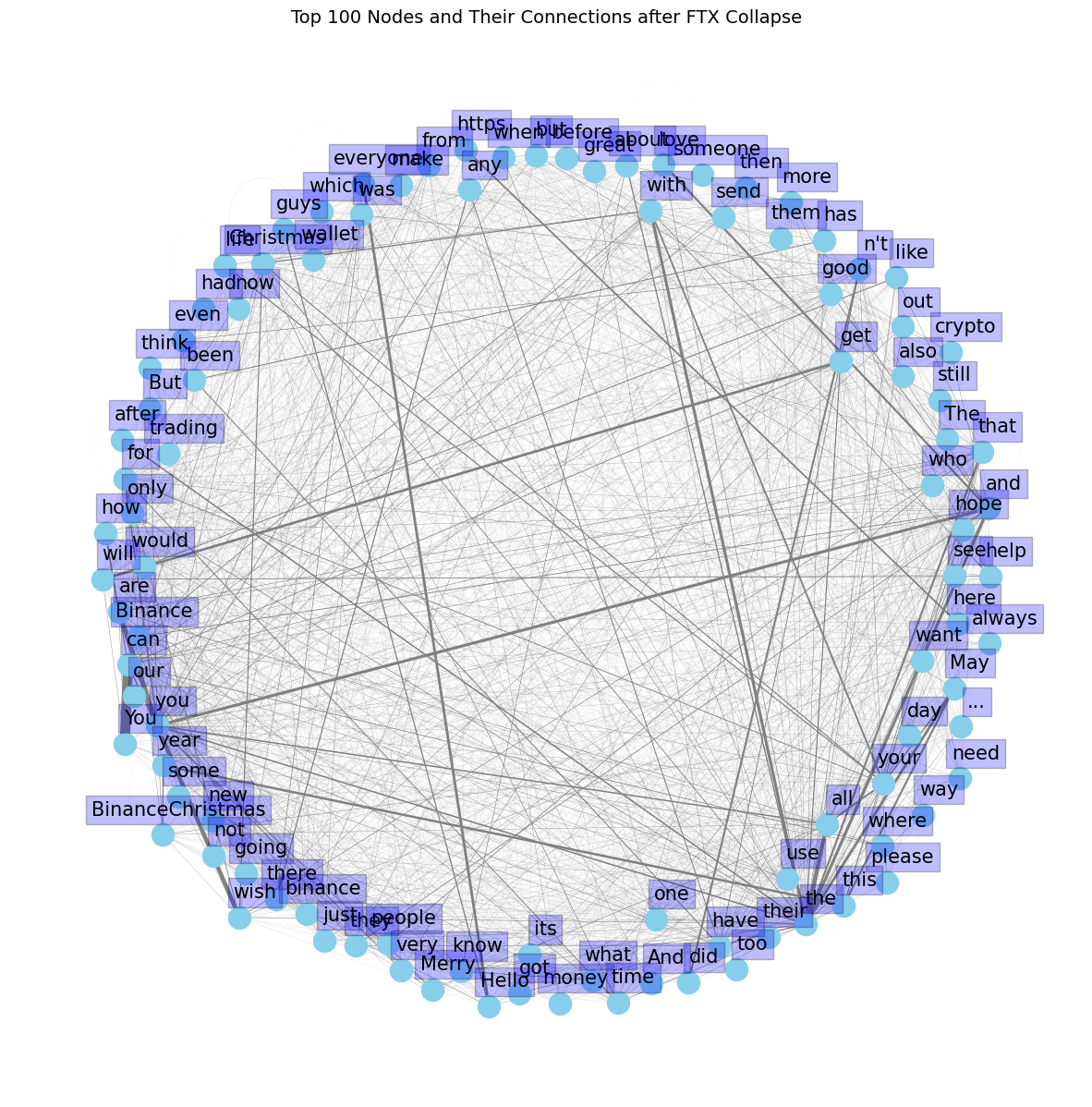}
    \caption{Network Analysis of Top 100 Nodes Post-FTX Collapse}
    \label{fig:net2}
\end{figure}

Network analysis quantifies structural changes in discourse through word co-occurrence networks (Figures \ref{fig:net1} and \ref{fig:net2}). Pre-collapse, the network exhibits \textbf{high edge density} with strong co-occurrence links connecting high-centrality nodes such as \textit{``trading,''} \textit{``wallet,''} \textit{``account,''} and \textit{``support''} (Fig. \ref{fig:net1}), indicating tightly-knit discussions centered on platform functionality. Post-collapse, \textbf{network density decreases markedly}; edges become sparser and centrality shifts toward seasonal terms (e.g., \textit{``Christmas,''} \textit{``merry''}) (Fig. \ref{fig:net2}). This topological shift---from cohesive functional discourse to fragmented, seasonally-dominated discussions---suggests the observed network dilution stems primarily from holiday topic intrusion rather than conclusive evidence of trust erosion, consistent with the non-significant sentiment trend in our RDD analysis.

\section{Conclusion and Future Directions}

This study uses the FTX collapse to examine trust dynamics in cryptocurrency markets, contributing mixed causal inference and NLP methods to the blockchain finance literature (see Appendix~\ref{app:related} for related work). Promising future directions include refined temporal controls, synthetic control methods, and domain-specific NLP models (see Appendix~\ref{app:conclusion_details} for detailed discussion). \footnote{Robustness analyses confirm the stability of our core findings: sensitivity tests employing alternative bandwidth specifications and estimation windows around the FTX collapse yield consistently significant results for both WETH price depreciation and NetFlow discontinuities, reinforcing the causal interpretation of institutional shocks on market behavior.}

\bibliographystyle{IEEEtran}
\bibliography{ref1}

\appendices

\section{Data Collection and Processing Details}
\label{app:data}
\begin{table*}[htbp]
\centering
\caption{Description of Variables and Empirical Ranges (January 2021 -- June 2023)}
\label{tab:data_description}
\resizebox{\textwidth}{!}{%
\begin{tabular}{@{}llclp{5.5cm}@{}}
\toprule
\textbf{Variable} & \textbf{Description} & \textbf{Type} & \textbf{Range (Min--Max)} & \textbf{Observation Period} \\
\midrule
\texttt{Gold\_price} & Gold spot price (USD/oz) & Numeric & 1,640 -- 2,050 & 01-Jan-2021 -- 30-Jun-2023 \\
\texttt{Price} & WETH average daily price (USD) & Numeric & 1,000.42 -- 4,752.88 & 01-Jan-2021 -- 30-Jun-2023 \\
\texttt{Netflow} & Net platform flow (DEX$-$CEX) in ETH & Numeric & $-6,110,470$ -- $+6,286,140$ & 01-Jan-2021 -- 30-Jun-2023 \\
\midrule
\multicolumn{5}{@{}l}{\textit{Raw Discord Data (Message-Level)}} \\
\texttt{AuthorID} & Discord user snowflake identifier & Numeric & $9.11 \times 10^{17}$ -- $1.02 \times 10^{18}$ & 01-Sep-2022 -- 27-Sep-2022 \\
\texttt{Author} & Discord username (including discriminator) & String & [Fugazi-lol\#9553, phoebe7200, \dots] & 01-Sep-2022 -- 27-Sep-2022 \\
\texttt{Timestamp} & Message timestamp (UTC-4) & DateTime & 2022-09-01 00:12 -- 2022-09-27 23:59 & 01-Sep-2022 -- 27-Sep-2022 \\
\texttt{Content} & Raw message text (Uniswap support queries) & Text & 0 -- 2,000 characters & 01-Sep-2022 -- 27-Sep-2022 \\
\texttt{Attachments} & Embedded media URLs (screenshots/tickets) & String & [Empty, discordapp.com/attachments/\dots] & 01-Sep-2022 -- 27-Sep-2022 \\
\texttt{Reactions} & Emoji reaction counts per message & String & [Null, \textbackslash U+2197\textbackslash U+FE0F (1), \dots] & 01-Sep-2022 -- 27-Sep-2022 \\
\midrule
\multicolumn{5}{@{}l}{\textit{Processed Sentiment Variables}} \\
\texttt{Sentiment\_b} & Binance Discord sentiment score & Numeric & $-0.504$ -- $+0.758$ & 24-Jul-2022 -- 30-Jun-2023 \\
\texttt{Label\_b} & Binance sentiment category (positive/neutral/negative) & Categorical & 47\% neutral, 30\% positive, 23\% negative & 24-Jul-2022 -- 30-Jun-2023 \\
\texttt{Sentiment\_u} & Uniswap Discord sentiment score (derived from \texttt{Content}) & Numeric & $-0.310$ -- $+0.364$ & 24-Jul-2022 -- 30-Jun-2023 \\
\texttt{Label\_u} & Uniswap sentiment category & Categorical & Predominantly neutral-to-positive & 24-Jul-2022 -- 30-Jun-2023 \\
\bottomrule
\end{tabular}%
}
\noindent\textbf{Note:} Financial market variables span the complete 912-day panel from January 2021 through June 2023. Sentiment variables commenced data collection on 24-July-2022 following the activation of Discord community monitoring, resulting in 342 daily observations (approximately 11.2 months) compared to 30.5 months for on-chain financial metrics. Raw Discord data represents the underlying message-level corpus ($n \approx 12,800$ messages) used to compute daily sentiment aggregates. The primary causal inference window (RDD/DID) focuses on September--December 2022 where all variables temporally converge.
\end{table*}

\vspace{0.5em}

Table~\ref{tab:data_description} presents the complete variable definitions, observation periods, and empirical ranges. Financial market variables span 01-Jan-2021 to 30-Jun-2023 ($n=912$ daily observations). Raw Discord message-level data and derived sentiment aggregates span 24-Jul-2022 to 30-Jun-2023 ($n=342$ daily observations), reflecting the later activation of community monitoring relative to on-chain data collection.

We selected Wrapped Ether (WETH) for our analysis due to its pivotal role in the Ethereum ecosystem and its widespread use in decentralized finance (DeFi). Unlike native ETH, WETH is an ERC-20 token, ensuring compatibility with other ERC-20 tokens and enhancing its usability and liquidity across both centralized (CEXs) and decentralized exchanges (DEXs) such as Uniswap and Balancer \citep{hagele2024centralized}. These consistent trading patterns allow for meaningful comparisons of transaction volumes and trust dynamics.

Our dataset comprises three integrated sources: (1) daily WETH prices and CEX/DEX transaction flows from Flipsidecrypto\footnote{https://flipsidecrypto.xyz/ }, spanning 01-Jan-2021 to 30-Jun-2023 and covering major platforms including Coinbase, OKX, Binance, Uniswap, and Sushiswap; (2) historical global gold prices from the World Gold Council\footnote{https://www.gold.org/ } as a stable control variable for macroeconomic trends; and (3) \textit{raw Discord server data} from Binance and Uniswap \textit{\#general} channels, extracted using DiscordChatExporter with message-level metadata fields (\textit{AuthorID}, \textit{Timestamp}, \textit{Content}, \textit{Attachments}, \textit{Reactions}), spanning 24-Jul-2022 to 30-Jun-2023.

\subsection{WETH Price and Transaction Flow}
The raw data originates from the Ethereum blockchain, queried via the Flipsidecrypto platform. Cross-validation confirmed data reliability: recent daily prices were verified against Ethereum queries, and five random transaction addresses were validated on Etherscan. The dataset covers major CEXs (Coinbase, LCX, BitBee, OKX, Binance) and DEXs (Uniswap, Sushiswap, Balancer, LuaSwap, SakeSwap, KyberSwap, 0x, ShibaSwap, Dydx, PancakeSwap, Solidly). As shown in Table~\ref{tab:data_description}, WETH prices range from \$1,000.42 to \$4,752.88 (01-Jan-2021 to 30-Jun-2023), while NetFlow (DEX inflow minus CEX inflow) ranges from $-6.1$M to $+6.3$M ETH.

\subsection{Global Gold Price}
Historical global gold prices (USD/oz) were obtained from the World Gold Council for the period 01-Jan-2021 to 30-Jun-2023 (range: \$1,640--\$2,050; see Table~\ref{tab:data_description}). Gold serves as a stable control variable to isolate crypto-specific effects from broader macroeconomic trends.

\subsection{Discord User Messages}
User messages from Discord's Binance and Uniswap communities were collected using DiscordChatExporter. Data collection activated on 24-Jul-2022 and continued through 30-Jun-2023, yielding approximately 12,800 message-level observations (Table~\ref{tab:data_description}) subsequently aggregated to 342 daily observations. This 11.2-month window captures the pre-FTX baseline (Jul--Nov 2022), the immediate collapse period (Nov 2022), and post-collapse recovery dynamics (Dec 2022--Jun 2023).

Raw message data were extracted with the following metadata structure (see Table~\ref{tab:data_description}):
\begin{itemize}
    \item \texttt{AuthorID}: Discord snowflake identifiers (numeric range: $9.11 \times 10^{17}$--$1.02 \times 10^{18}$) serving as anonymized user tokens;
    \item \texttt{Author}: Discord usernames including discriminators (e.g., \texttt{Fugazi-lol\#9553}, \texttt{phoebe7200});
    \item \texttt{Timestamp}: Message timestamps in ISO 8601 format (UTC-4), enabling temporal aggregation;
    \item \texttt{Content}: Raw message text (0--2,000 characters) containing support queries, trading discussions, and community interactions;
    \item \texttt{Attachments}: URLs to embedded media (screenshots, transaction logs, support tickets);
    \item \texttt{Reactions}: Emoji reaction counts indicating community engagement intensity.
\end{itemize}

Messages were extracted primarily from public \#general and \#help channels where user sentiment regarding platform stability and trading conditions is most prevalent. Author IDs were treated as randomly generated numeric codes for longitudinal message matching without identifying specific users. The anonymization process adhered to ethical and privacy guidelines with no confidential data compromised; original usernames were retained only in the secure raw dataset for verification purposes while analytical datasets utilized hashed identifiers.

\subsection{Sentiment Scoring Methodology}
For sentiment analysis, we employed the \textit{twitter-roberta-base-sentiment-latest} model from Hugging Face\footnote{https://huggingface.co/cardiffnlp/twitter-roberta-base-sentiment-latest }, a fine-tuned RoBERTa model optimized for social media linguistic nuances. We classified messages into \textit{positive} (1), \textit{negative} ($-1$), or \textit{neutral} (0) categories and calculated daily average sentiment scores as the weighted average of these values.

Sentiment analysis employed the cardiffnlp/twitter-roberta-base-sentiment-latest model from Hugging Face, a RoBERTa-based LLM fine-tuned on Twitter data. This model was selected for its optimization for social media linguistic nuances prevalent in Discord communities (e.g., emoji usage, crypto-specific terminology).

The processing pipeline aggregated raw \texttt{Content} fields to daily sentiment metrics:
\begin{enumerate}
    \item \textit{Message-level scoring}: Each \texttt{Content} text was processed individually, generating continuous sentiment scores and categorical labels (positive, neutral, negative);
    \item \textit{Daily aggregation}: Message-level scores were averaged by date (based on \texttt{Timestamp}) to create daily sentiment indices;
    \item \textit{Community separation}: Binance (CEX) and Uniswap (DEX) communities were processed separately to maintain platform-specific signals.
\end{enumerate}

As documented in Table~\ref{tab:data_description}, Binance sentiment scores range from $-0.504$ to $+0.758$ with distribution 47\% neutral, 30\% positive, 23\% negative; Uniswap sentiment ranges from $-0.310$ to $+0.364$, exhibiting predominantly neutral-to-positive valence. The narrower Uniswap range reflects the relative stability of DEX communities compared to CEX volatility during the FTX crisis period. The raw message corpus ($n \approx 12,800$) underlying these aggregates enables robust daily estimation even during high-volatility periods when message volume spikes.

\section{Methodological Details}
\label{app:method}

\subsection{Regression Discontinuity Design (RDD)}
To examine the causal effects at the event threshold (November 11, 2022), we estimate:
\begin{dmath}
    Y_{it} = \alpha + \beta D_t + f(X_{it}) + \epsilon_{it}
\end{dmath}
where $Y_{it}$ represents the outcome variable (WETH price, NetFlow, or sentiment score); $D_t$ is an indicator variable implying whether the observation occurs before or after the event ($D_t = 1$ if $t \geq T_0$); $f(X_{it})$ captures the smooth relationship between the outcome and the running variable (e.g., time); and $\beta$ quantifies the discontinuity at the threshold, representing the causal impact of the event. We chose two months before and after FTX collapse as our analysis time.

\subsection{Difference-in-Differences (DID)}
To further validate the impact on WETH prices using gold prices as a control to account for broader market trends, we estimate:
\begin{dmath}
    Y_{it} = \alpha + \gamma Post_t + \delta Treatment_i + \theta (Post_t \cdot Treatment_i) + \epsilon_{it}
\end{dmath}
where $Y_{it}$ is the observed WETH price, $Post_t$ indicates the post-event period ($Post_t = 1$ if $t \geq T_0$), and $Treatment_i$ identifies whether the observation belongs to the treatment group (WETH). The interaction term $(Post_t \cdot Treatment_i)$ captures the differential effect of the event on WETH prices relative to the control group (gold). The coefficient $\theta$ represents the causal impact of the event on WETH prices after accounting for market-wide effects.

\subsection{Latent Dirichlet Allocation (LDA)}
We applied LDA, a generative probabilistic model, to uncover latent topics and semantic structures in user discussions. The joint probability of observing a word is given by:
\begin{dmath}
    P(w | \alpha, \beta) = \prod_{d=1}^{D} \int P(\theta_d | \alpha) \prod_{n=1}^{N_d}
    \sum_{z=1}^{K} P(z_n | \theta_d) P(w_n | \phi_z) \, d\theta_d
\end{dmath}
The model outputs a set of topics, where each topic is represented as a probability distribution over words, and each document is assigned a mixture of these topics. We selected $K=8$ topics based on coherence scores (Figure \ref{fig:co_combine}); optimization details appear in Appendix \ref{app:method}.

\begin{figure}[ht!]
    \centering
    \includegraphics[width=0.9\columnwidth]{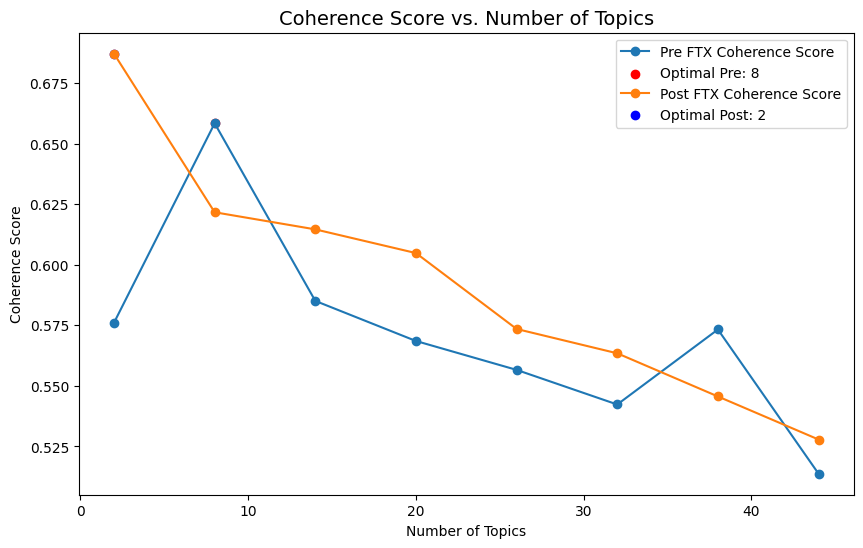}
    \caption{Coherence Score vs. Number of Topics Before and After Event}
    \label{fig:co_combine}
\end{figure}

\subsection{Network Analysis}
Network analysis was applied to visualize and evaluate the connections between words and topics, providing additional insights into the structure and focus of user discussions. We constructed word co-occurrence networks where nodes represent words and edges represent the co-occurrence frequency of words within a predefined sliding window. The strength of an edge between words $w_i$ and $w_j$ is quantified by the normalized co-occurrence weight $\omega_{ij}$, defined as:
\begin{equation}
    \omega_{ij} = \frac{\chi_{ij}}{\mathcal{T}}
    \label{eq:edgeweight}
\end{equation}
where $\chi_{ij}$ denotes the raw co-occurrence frequency of words $w_i$ and $w_j$ within the sliding window, and $\mathcal{T}$ represents the total co-occurrence frequency across all word pairs in the corpus.

Pre- and post-event networks were analyzed using metrics such as edge density, which measures the overall connectivity of the network, and node centrality, which identifies critical terms dominating discussions.

\subsection{LDA Coherence Optimization}
The number of topics, $K$, is determined using the coherence score, a metric that evaluates the semantic similarity between high-probability words within a topic. Before the FTX collapse, the coherence score peaked at $K=8$, indicating strong internal consistency within topics. After the collapse, the coherence score peaked at $K=2$, suggesting a shift towards more concentrated discussions. However, $K=2$ oversimplifies the diversity of topics, and $K=8$ represented the second-best coherence score post-event, with values similar to those before the event. For consistency and interpretability, we selected $K=8$ for both periods, allowing meaningful comparisons of topic distributions.

\subsection{Network Construction and Analysis}
Network analysis was applied to visualize and evaluate the connections between words and topics, providing additional insights into the structure and focus of user discussions. We constructed word co-occurrence networks where nodes represent words and edges represent the co-occurrence frequency of words within a predefined sliding window.

Pre- and post-event networks were analyzed using metrics such as edge density, which measures the overall connectivity of the network, and node centrality, which identifies critical terms dominating discussions. Before the FTX collapse, networks displayed high connectivity, reflecting focused and cohesive discussions centered on trading and platform functionalities. After the event, networks became sparser and less connected, indicating a diversification of discussion topics. These changes highlight a shift in user focus from transactional issues to concerns about platform security, governance, and transparency.

\section{Detailed Results and Topic Analysis}
\label{app:results}
\subsection{Regression Discontinuity Design Estimates}
Table~\ref{tab:rdd_results} reports the complete causal estimates for the FTX collapse impact across price, NetFlow, and sentiment metrics. The estimates confirm significant discontinuities in WETH pricing and fund migration patterns, while sentiment measures exhibit substantial variance as reflected in the wide confidence intervals for both exchange communities.
\begin{table}[htbp]
\caption{Regression Discontinuity Design Estimates}
\label{tab:rdd_results}
\centering
\small % Scale text slightly to accommodate wide CIs without resizebox
\begin{tabular}{@{}l r l@{}}
\toprule
\textbf{Outcome} & \textbf{Estimate} & \textbf{95\% Confidence Interval} \\
\midrule
WETH Price (USD) & $-223.82$ & $[-223.92, -223.72]$ \\
NetFlow (USD) & $-291{,}928.65$ & $[-291{,}928.75, -291{,}928.55]$ \\
Sentiment (Binance) & $0.02$ & $[-0.08, 0.12]$ \\
Sentiment (Uniswap) & $-0.05$ & $[-0.15, 0.05]$ \\
\bottomrule
\end{tabular}
\end{table}

\subsection{Latent Dirichlet Allocation Topic Terms}
Table~\ref{tab:topic_words} details the top-10 most salient terms for each of the eight topics identified via LDA, comparing pre- and post-collapse discourse. Notable semantic shifts include Topic~6's transition from functional trading vocabulary (\textit{spot}, \textit{wallet}, \textit{bnb}) to holiday-related expressions (\textit{christmas}, \textit{merry}, \textit{wish}), and Topic~3's emergence of security-focused terms (\textit{scams}, \textit{scammers}, \textit{report}) post-event. These lexical changes corroborate the sentiment analysis findings discussed in Section~V, explaining the apparent contradiction between quantitative sentiment stability and underlying trust erosion in the Binance community.

% In document:
\begin{table}[htbp]
\caption{Comparison of Top-10 Terms per Topic Pre- and Post-FTX Collapse}
\label{tab:topic_words}
\centering
\footnotesize % Smaller font to prevent overflow
\setlength{\tabcolsep}{3pt} % Tighten column padding
\begin{tabular}{@{}l>{\raggedright\arraybackslash}p{3.5cm}>{\raggedright\arraybackslash}p{3.5cm}@{}}
\toprule
\textbf{Topic} & \textbf{Pre-FTX Collapse} & \textbf{Post-FTX Collapse} \\
\midrule
Topic 1 & think, like, trade, yeah, binance, got, just, know, don, hello & just, really, think, got, lol, yeah, time, don, thanks, know \\
\addlinespace[2pt]
Topic 2 & dm, server, just, discord, check, ok, link, binance, guys, thanks & did, trade, just, bro, ur, money, gift, like, thank, hey \\
\addlinespace[2pt]
Topic 3 & know, nft, problem, buy, use, make, crypto, money, account, binance & group, right, nice, mean, telegram, people, gm, scams, scammers, report \\
\addlinespace[2pt]
Topic 4 & contact, answer, bot, chat, www, en, support, https, com, binance & binancians, day, ok, doing, evening, ex\_binance, guys, morning, hi, good \\
\addlinespace[2pt]
Topic 5 & deleted, love, did, inquiry, community, channel, ask, question, need, help & market, buy, post, inquiry, question, channel, community, ask, need, help \\
\addlinespace[2pt]
Topic 6 & right, dont, doing, spot, address, lol, bnb, thank, yes, wallet & hope, binance, love, wish, happy, new, year, merry, binancechristmas, christmas \\
\addlinespace[2pt]
Topic 7 & favorite, bro, report, binanceblockchainweek, yo, moment, welcome, gm, scams, scammers & discord, admin, community, best, dm, scam, binance, just, yes, hello \\
\addlinespace[2pt]
Topic 8 & crypto, earn, support, announcement, www, en, com, https, hi, binance & crypto, use, wallet, nft, www, en, support, https, com, binance \\
\bottomrule
\end{tabular}
\end{table}

\section{Related Work}
\label{app:related}
This study bridges behavioral finance, blockchain analytics, and Trust Theory, integrating causal inference with NLP to examine trust dynamics across exchange platforms.

\subsection{Behavioral Finance and Trust}
Trust---defined as the subjective probability that an agent will fulfill obligations \citep{starr2011trust}---facilitates transactions and influences asset pricing, though real-world decisions systematically deviate from efficient market hypotheses due to risk perception biases \citep{barberis2018psychology, ricciardi2008psychology}. Psychological mechanisms including extrapolation, overconfidence, and sticky beliefs explain how investors form expectations under uncertainty \citep{barberis2018psychology}. While prior work establishes that trust influences crypto market development \citep{park2022impact, rehman_salah_damiani_svetinovic_2019, arli_van_esch_bakpayev_laurence_2020, marella_upreti_merikivi_tuunainen_2020}, we advance this literature by quantifying the immediate behavioral impact of systemic CEX failure through measurable fund flows between centralized and decentralized exchanges.

\subsection{Sentiment Analysis in Finance and Blockchain}
Financial community sentiment correlates with market performance \citep{7381513}, with applications spanning P2P lending trust dynamics \citep{Niu2020}, blockchain governance forums \citep{10727009}, and transaction prediction \citep{Yuan2021}. We extend RoBERTa-based sentiment analysis to exchange-specific Discord communities (Binance, Uniswap), capturing granular trust dynamics during the FTX collapse.

\subsection{Machine Learning and Causal Inference in Blockchain}
AI integration spans blockchain infrastructure \citep{blockchain_security_enhancement, Islam_2024, ElHessaini2024}, cross-chain applications \citep{9274451, Yu2022}, and Web3 systems \citep{zhang2023machine}. Causal inference methods evaluate blockchain policies: Chen et al. apply staggered DID to contractual incompleteness \citep{RePEc:inm:ormnsc:v:69:y:2023:i:11:p:6540-6567}; Wang et al. employ triple-difference (DDD) for green innovation analysis \citep{Wang2023}; Liu et al. use RDD to analyze EIP-1559 fee mechanisms \citep{10.1145/3548606.3559341}. We contribute by applying RDD and DID to quantify trust migration effects, using gold prices as a control to isolate crypto-specific impacts of the FTX collapse on WETH valuation and exchange flows.

\section{Intellectual Merits, Practical Impacts, and Future Research}
\label{app:conclusion_details}
\subsection{Intellectual Merits and Practical Impacts}
Intellectually, this research bridges behavioral finance with blockchain analytics to examine trust dynamics across exchange platforms, quantifying how institutional shocks trigger capital reallocation and behavioral responses in digital asset markets. We combine causal inference (RDD/DID) with on-chain analytics and NLP to operationalize sociotechnical trust---a dynamic process requiring continuous alignment between technical systems and user communities \citep{10.1145/3789256}. Unlike descriptive event studies, we isolate causal effects of systemic disruptions by exploiting blockchain's inherent transparency: the NetFlow discontinuity represents verifiable on-chain behavioral change, using fund flows to proxy preference shifts when direct trust measurement is impossible. Topic modeling and network analysis further reveal structural discourse evolution that price data alone cannot capture, bridging technical blockchain research with computational social science.

Practically, our findings underscore the fragility of centralized trust architectures and the resilience of decentralized alternatives during systemic crises, offering critical insights for exchange operators regarding transparency requirements and for regulators assessing risk contagion across platform types. We advance open science by releasing analytical code and curated datasets to facilitate reproducibility and extension across disciplinary boundaries---enabling finance scholars to leverage our causal inference frameworks, computer scientists to adapt our NLP pipelines, and sociologists to utilize our community discourse corpora for studying digital trust formation.

\subsection{Limitations and Future Research}
Our analysis opens three promising avenues for future inquiry. First, seasonal confounds introduced by the holiday season coincidence warrant refined temporal controls; future work should employ geospatial filtering to isolate less holiday-influenced regions \citep{chen2024global} or pioneer statistical holiday adjustments, while employing multi-token indices to validate systematic claims across diverse market conditions.

Second, future research can enhance identification precision through synthetic control methods \citep{abadie2005semiparametric}. While gold capably captures broad macroeconomic trends, augmenting it with optimally weighted combinations of additional safe-haven assets---sovereign bonds, inflation-protected securities, and commodities---would further distinguish exchange-specific trust collapses from general flight-to-safety behavior.

Third, methodological advances in NLP and temporal modeling remain essential. Advancing beyond our Twitter-trained RoBERTa model, future work can develop domain-specific sentiment classifiers tailored to Discord-specific crypto vernacular (slang, sarcasm), or employ LLM-based denoising encoders to improve the extraction of trust-relevant signals. Applying advanced temporal models such as CausalStock \citep{li2024causalstock} will further clarify whether observed migrations represent temporary reactions or permanent structural transitions in decentralized trust architectures.

\end{document}